\documentclass[a4paper,10pt]{article}
\usepackage{graphicx}
\usepackage{epstopdf}
\usepackage{url}
\usepackage{cite}
\usepackage{algorithmicx}
\usepackage[ruled]{algorithm}
\usepackage{algpseudocode}
\usepackage{amssymb}
\usepackage{amsmath}
\usepackage{color}
\usepackage{picins} 
\usepackage{fullpage}
\usepackage{authblk}
  
\newcommand{\revs}[1]{\textcolor{black}{#1}}

\begin{document}

\title{Control Theoretic Optimization of 802.11 WLANs: Implementation and Experimental Evaluation}

\author[1]{Pablo Serrano}
\author[2]{Paul Patras}
\author[3,1]{Andrea Mannocci}
\author[3,1]{Vincenzo Mancuso}
\author[1,3]{Albert Banchs}

\affil[1]{University Carlos III de Madrid, Spain}
\affil[2]{Hamilton Institute, Ireland}
\affil[3]{Institute IMDEA Networks, Spain}

\date{}

\maketitle

\begin{abstract}
In 802.11 WLANs, adapting the contention parameters to network conditions results in substantial performance improvements. Even though the ability to change these parameters has been available in standard devices for years, so far no adaptive mechanism using this functionality has been validated in a realistic deployment. In this paper we report our experiences with implementing and evaluating two adaptive algorithms based on control theory, one centralized and one distributed, in a large-scale testbed consisting of 18 commercial off-the-shelf devices. We conduct extensive measurements, considering different network conditions in terms of number of active nodes, link \revs{qualities, and data traffic}. We show that both algorithms significantly outperform the standard configuration in terms of total throughput. We also identify the limitations inherent in distributed schemes, and demonstrate that the centralized approach substantially improves performance under a large variety of scenarios, which confirms its suitability for real deployments.
\end{abstract}

\section{Introduction}
The IEEE 802.11 standard for Wireless LANs~\cite{80211revised} has become one of the most commonly used technologies to provide broadband connectivity to the Internet. The default channel access mechanism employed in IEEE 802.11 networks is based on a CSMA/CA scheme, regulated by a set of parameters that determines the aggressiveness of the stations when trying to access the channel. In particular, the contention window ($CW$) parameter controls the probability that a station defers or transmits a frame once the medium has become idle, and therefore has a key impact on the WLAN performance. 

Commercial devices implement a fixed $CW$ configuration, which is known to yield suboptimal performance \revs{\cite{bianchi00}}. Indeed, for a fixed $CW$, if too many stations contend the collision rate will be very high, while if few stations are backlogged the channel will be underutilized most of the time. This behavior has been analyzed by several works in the literature, e.g. \revs{\cite{bianchi00,comnet,tvt}}, which have shown that adapting the $CW$ to the number of backlogged stations significantly improves performance.

Following the above result, an overwhelming number of solutions have proposed to adapt the 802.11 MAC behavior to the observed network conditions with the goal of maximizing the WLAN performance \cite{freitag06,scalia06,nafaa05,yang07,xia06,ni03,slow05,slow10,heusse05,siris06, AOBimpl}. 
However, as we detail in the related work section, these previous works suffer from both theoretical and practical limitations.

In this paper, we present our experiences with the implementation of two adaptive algorithms, namely the \emph{Centralized Adaptive Control} (CAC) \cite{patras09monet} and the \emph{Distributed Adaptive Control} (DAC) \cite{patras10tmc}, both based on a Proportional Integrator (PI) controller that dynamically tunes the $CW$ configuration to optimize performance. In contrast to previous proposals \revs{\cite{freitag06,scalia06,nafaa05,yang07,xia06,ni03,slow05,slow10,heusse05,idle07, siris06, AOBimpl}}, both algorithms are supported by solid theoretical foundations from control theory and can be easily implemented with unmodified existing devices. \revs{Further, as compared to the few existing implementations of adaptive MAC mechanisms that require complete interface redesign and enhanced computational capabilities \cite{AOBimpl} or rely on proprietary firmware code and are tight to a specific platform \cite{idle07}, our prototypes can be deployed with off-the-shelf devices, demand minimal system resources and do not alter the networking stack as they are executed as userland processes.}

\revs{CAC and DAC have been carefully designed following detailed analyses of the 802.11 performance and employing control-theoretic techniques, to ensure both optimal performance and system stability, while previous simulations illustrate the potential benefits of these algorithms \cite{patras09monet,patras10tmc}. In this paper, we take our earlier work one step further and demonstrate the feasibility of running these mechanisms on real devices. Further, we assess their behavior extensively in a real 802.11 deployment under a wide range of network conditions.}

First, we provide a detailed description of the implementation of our adaptive mechanisms with commodity hardware and open-source drivers. The algorithms run as user space applications and rely on standardized system calls to estimate the contention level in the WLAN and adjust the $CW$ configuration of 802.11 stations. 
We also provide insights into the differences between the theoretical design and the practical implementation of the algorithms, which arose with the inherent limitations of the real devices. 
Second, by conducting exhaustive experiments in a large-scale testbed consisting of 18 devices, \revs{that captures the specifics of realistic deployments such as office environment, seminar room, hotel lobby, etc.,} we evaluate the performance of our proposals under non-ideal channel effects and different traffic conditions. Additionally, we compare the performance of our algorithms against the default IEEE 802.11 configuration, and identify those scenarios where a network deployment can benefit from using such adaptive mechanisms.

Our results confirm that both approaches outperform the standard's default scheme, improving the performance by up to 50\%. Our experiments also reveal that the distributed algorithm suffers from a number of problems with heterogeneous radio links, which are inherent in its distributed nature and the limitations of the wireless interfaces. In contrast, the centralized scheme exhibits remarkable performance under a wide variety of network conditions. The conclusions drawn from our analysis prove the feasibility of using adaptive MAC mechanisms in realistic scenarios and provide valuable insights for their design.

The remainder of the paper is organized as follows. Section~\ref{sec:background} summarizes the IEEE 802.11 EDCA protocol and the underlying principles of CAC and DAC. Section~\ref{sec:implementation} details the implementation of the functionality comprised by the proposed schemes. Section~\ref{sec:testbed} describes our testbed and the validation of the implementation of the algorithms. Section~\ref{sec:performance} presents a thorough experimental study of the algorithms in a wide set of network conditions.  \revs{Section~\ref{sec:related} summarizes the related work. Finally, Section~\ref{sec:conclusions} concludes the paper.}

\section{Background}
\label{sec:background}

This section summarizes the behavior of IEEE 802.11 EDCA \revs{\cite{80211revised}} and the two adaptive protocols implemented in this paper. 

\subsection{IEEE 802.11 EDCA}

The IEEE 802.11 Enhanced Distributed Channel Access (EDCA) mechanism \revs{\cite{80211revised}} is a CSMA/CA-based protocol that operates as follows. If a station with a new frame to transmit senses the channel idle for a period of time equal to the arbitration interframe space parameter ($AIFS$), the station transmits. Otherwise, if the channel is busy (either immediately or during the $AIFS$ period), the station continues to monitor the channel until it is sensed idle for an $AIFS$ interval, and then executes a backoff process.

\begin{figure}
\centerline{\includegraphics[width=\columnwidth]{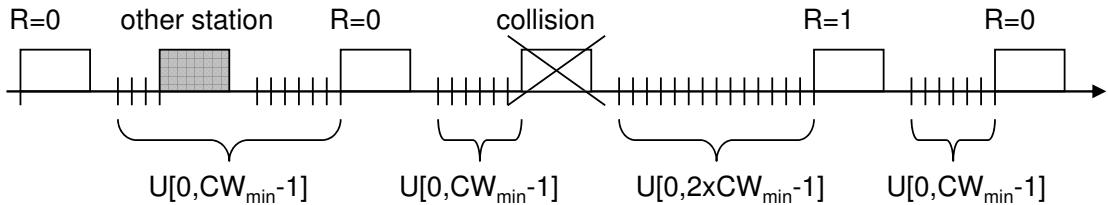}}
\caption{Retry flag marking upon collisions.}
\label{fig:retry}
\end{figure}

Upon starting the backoff process, stations compute a random integer uniformly distributed in the range $[0,CW-1]$, and initialize their backoff time counter with this value. The $CW$ value is called the contention window, and depends on the number of failed transmission attempts. For the first transmission attempt the minimum contention window ($CW_{min}$) is used. In case of a collision, its value doubles, up to a maximum value $CW_{max}$. The backoff time counter is decremented once every time slot if the channel is sensed idle, frozen when a transmission is detected on the channel, and reactivated when the channel is sensed idle again for an $AIFS$ time. When the backoff time counter reaches zero, the station transmits its frame in the next time slot. 

When two or more stations start transmitting simultaneously, a collision occurs. Acknowledgment (ACK) frames are used to notify a transmitting station of successfully received frames. In the case of a failed transmission, the station doubles its $CW$ and reenters the backoff process. Once a frame has been successfully transmitted or the retry limit has been exceeded, the $CW$ value is set again to $CW_{min}$. To prevent duplicates, the standard uses a retry flag $R$ to mark those frames that are being retransmitted, i.e., the flag is set to 0 on the first transmission attempt, and set to 1 on every retransmission (see Fig.~\ref{fig:retry}). As we discuss later, our algorithms exploit this functionality to infer the network conditions and adapt the $CW$ of the stations accordingly. 

To support service differentiation, EDCA implements different access categories (ACs) at every station, each having a different backoff configuration. The parameters of each AC are announced  by the Access Point using the Beacon Frames. In the rest of the paper we do not consider service differentiation and assume that all stations only execute the Best Effort AC. 

\subsection{Optimal Point of Operation of the WLAN}
\label{sec:optimal_point}

Both CAC and DAC share the goal of adjusting the $CW$ to drive the WLAN to the optimal point of operation that maximizes the total throughput given the observed network conditions. Let $p$ denote the probability that a transmission attempt collides. \revs{Following the work of \cite{bianchi00}, that derives the optimal transmission probability $\tau_{opt}$, the corresponding optimal conditional collision probability can be computed as
\begin{equation}
p_{col} = 1 - (1-\tau_{opt})^{n-1}= 1 - \left(1-\frac{1}{n} \sqrt{\frac{2 T_e}{T_c}}\right)^{n-1}. 
\end{equation}
We have shown in \cite{patras09monet,patras10tmc} that for large $n$ values the above can be approximated by}
\begin{equation}
p_{opt} \approx 1 - e^{-\sqrt{\frac{2T_e}{T_c}}},
\label{eq:p_opt}
\end{equation}
where $T_e$ is the duration of an idle slot (a PHY layer constant) and $T_c$ is the average duration of a collision. \revs{Further, previous simulations demonstrate that this approximation does not impact the performance even when $n$ is small \cite{patras09monet,patras10tmc}.} Therefore, $p_{opt}$ does not depend on the number of stations, but only on the average duration of a collision $T_c$, i.e., 
$$
T_c = T_{PLCP} + \frac{E[L]}{C} + EIFS.
$$
where $T_{PLCP}$ is the duration of the Physical Layer Convergence Protocol (PLCP) preamble and header, $C$ is the modulation rate and $EIFS$ is a PHY layer constant. \revs{$E[L]$ denotes the expected length of the longest packet involved in a collision, which, by ordering stations according to the length of their transmitted packets, $L_i$, can be computed as
$$
E[P] = \frac{1}{P_c}\sum_{i=1}^n \tau \left(1-(1-\tau)^{i-1}\right)(1-\tau)^{n-i-1}L_i,
$$
where $P_c$ is the average probability that a slot contains a collision.}

\subsection{Centralized Adaptive Control Algorithm}

\begin{figure}
\centerline{\includegraphics[width=\columnwidth]{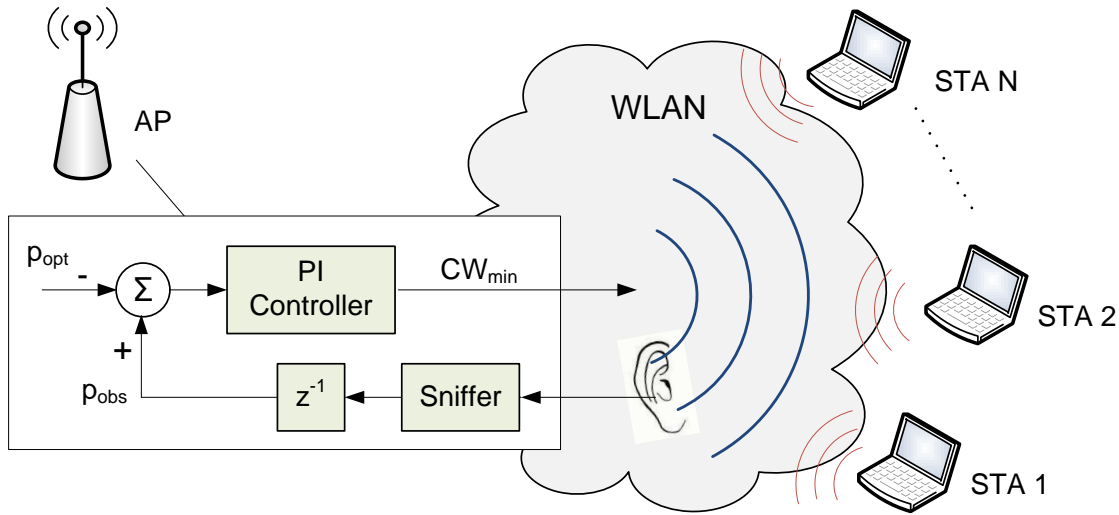}}
\caption{CAC algorithm.}
\label{fig:cac}
\end{figure}

The \emph{Centralized Adaptive Control} (CAC) algorithm \cite{patras09monet}, illustrated in Fig.~\ref{fig:cac}, is based on a PI controller located at the Access Point (AP). This controller computes the configuration of the $CW_{min}$ parameter as an integer ranging between the default minimum and maximum values defined by the standard specification, while $CW_{max}$ is set as $CW_{max} = 2^m CW_{min}$, following the standard binary exponential backoff procedure.\footnote{In our experiments we use the PHY layer parameters of IEEE 802.11a, hence $m = 6$ is chosen.} \revs{By adjusting the $CW$  configuration to be used by the stations, the AP controls the collision probability in the WLAN, with the goal of driving the network to the optimal point of operation that maximizes throughput.}

Following the above, the controller performs two tasks every beacon interval (approx. 100~ms)\revs{\footnote{This is the default beacon interval value recommended by the 802.11 specification and standard compliant devices must be able to update their configuration with this frequency. Setting this parameter to a larger value is permitted by current devices, however larger values may slow the reaction to network changes.}}: ($i$)~it estimates the current point of operation of the WLAN as given by the observed collision probability $p_{obs}$, and ($ii$)~based on this estimation and $p_{opt}$, it computes the $CW$ configuration to be used during the next beacon interval and sends it to the stations in a beacon frame.

The computation of $p_{obs}$ is based on the observation of the retry flag of successful frames. Let us denote by $R_1$ ($R_0$) the number of observed frames with the retry bit set (unset) during a beacon interval. Assuming that no frames exceed the retry limit given by the {\ttfamily MAX\_RETRY} parameter,\footnote{Note that this assumption is accurate as in an optimally configured WLAN the collision probability is very low.} and that \revs{transmission attempts} collide with a constant and independent probability,\footnote{This assumption has been widely used and shown to be accurate, see e.g. \cite{bianchi00}.} the observed probability of a collision in the WLAN can be estimated with~(see \cite{patras09monet}):
\begin{equation}
\label{eq:p_others}
p_{obs} = \frac{R_1}{R_0+R_1}.
\end{equation}
\revs{which can be seen as the probability that the first transmission attempt from a station collides. Given the short duration of a frame transmission relative to the beacon interval we expect the above estimation to yield good accuracy. However, to avoid that a too small number of samples induces a high degree of inaccuracy in this estimation, if the number of samples used to compute $p_{obs}$ is smaller than 20, the update is  deferred until the next beacon.\footnote{With an expected average $p_{obs} \approx 0.16$ and $N = 20$ samples, the size of the 95\% confidence interval will be at most $\pm 0.16$, which is acceptable.}}

\revs{To calculate the new $CW_{min}$, CAC employs a PI controller that takes as input an error signal $e$, computed as the difference between the observed collision probability $p_{obs}$ and the target value $p_{opt}$:}
\begin{equation}
\label{eq:e_cac}
e = p_{obs} - p_{opt}.
\end{equation}

In this way, when the observed collision probability is above the target value, the error signal will be positive and trigger an increase of the $CW_{min}$, and consequently a decrease of the collision rate in the next beacon interval. Similarly, when the collision probability is below the target value, $CW_{min}$ is decreased in order to increase the activity on the channel. \revs{The operation of CAC is summarized in Algorithm~\ref{alg:cac}.}

\begin{algorithm}[!t]
 \caption{Centralized Adaptive Control algorithm.}
 \label{alg:cac}
 \begin{algorithmic}[1]
	\While{true}
		\Repeat
			\If{new frame sniffed}
					\State retrieve retry flag			
					\If {retry flag is set}
						\State Increment $R_1$
					\Else
						\State Increment $R_0$
					\EndIf
			\EndIf
		\Until {new beacon interval}
		\State compute $p_{obs}[t]$ using (\ref{eq:p_others}) 	
		\State $e[t] = p_{obs}[t]-p_{opt}$	
		\State $CW_{min}[t] = CW_{min}[t - 1] + K_P \cdot e[t] + $
		\State \hspace{4.2em} $ + (K_I - K_P) \cdot e[t - 1]$
			
		\State send beacon with new $CW$ configuration
	\EndWhile
\end{algorithmic}
\end{algorithm}

The $\{K_P,K_I\}$ parameters of the PI controller are obtained using the Ziegler-Nichols rules, to achieve a proper trade-off between stability and speed of reaction to changes.
\revs{Specifically, these are computed as $K_p = 0.4 K_u$, respectively $K_i = K_p/(0.85T_i)$, where $K_u$ is the $K_p$ value that turns the system unstable when $K_i = 0$ and $T_i$ is the oscillation period under these conditions \cite{franklin}.} Then, $K_p$ and $K_i$ are configured as follows:

\begin{equation}
\left.
\right.\begin{aligned}
K_P &= \frac{0.8}{p_{opt}^2 (1 + p_{opt}\sum_{k = 0}^{m-1}{(2p_{opt})^k})}; \\
K_I &= \frac{0.4}{0.85 \cdot p_{opt}^2 (1 + p_{opt}\sum_{k = 0}^{m-1}{(2p_{opt})^k})}.
\end{aligned}
\label{eq:ctrl_params}
\vspace*{1em}
\end{equation}
\revs{The detailed computation of the  $\{K_P,K_I\}$ parameters, as well as the formal proof that the system behaves stably with this configuration are given in \cite{patras09monet}.}

\subsection{Distributed Adaptive Control Algorithm}

\begin{figure}%
\includegraphics[width=0.9\columnwidth]{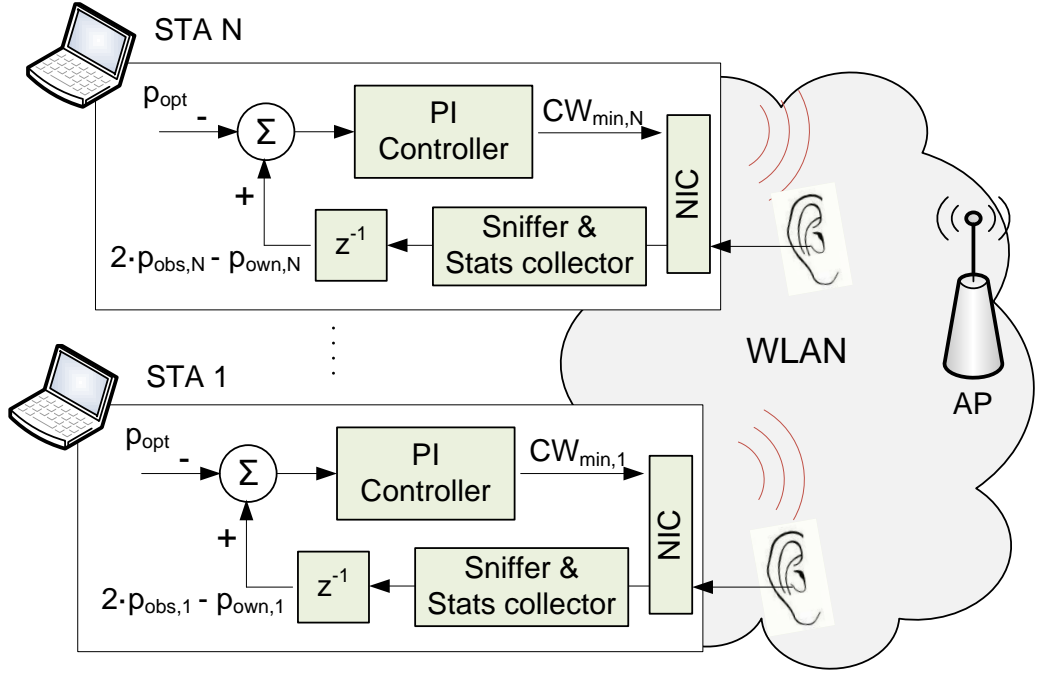}%
\caption{DAC algorithm.}%
\label{fig:dac}%
\end{figure}

The \emph{Distributed Adaptive Control} (DAC) algorithm \cite{patras10tmc} employs an independent PI controller at each station to compute its $CW$ configuration, to drive the overall collision probability to the target value $p_{opt}$. As illustrated in Fig.~\ref{fig:dac}, each controller computes the $CW_{min}$ value employed by its Network Interface Card (NIC), based on the locally observed network conditions. Similarly to CAC, $CW_{max}$ is set as $CW_{max} = 2^m CW_{min}$.

While with centralized approaches all stations use the same configuration provided by a single entity, and therefore fairly share the channel, with distributed approaches this is not necessarily the case. \revs{While ensuring the collision probability in the WLAN is consistent with the optimal value derived in Sec.~\ref{sec:optimal_point} guarantees the maximization of the total network throughput, due to the decentralized fashion in which DAC operates, this objective can be achieved by having stations configured differently. More precisely, each node~$i$ is responsible for its own configuration and transmits with an independent probability $\tau_i$. Since the collision probability in a network of $n$ stations is given by $p=1-\prod_{i=1}^n(1-\tau_i)$, it is easy to observe that there may exist multiple $(\tau_1,...,\tau_n)$ solutions that satisfy $p=p_{opt}$. However, this dissimilar $CW$ configuration will likely yield significant unfairness among stations. Therefore,} to guarantee a fair throughput distribution, the error signal utilized in DAC consists of two terms: one to drive the WLAN to the desired point of operation, and another one to achieve fairness between stations. More specifically, the error signal at station $i$ is given by
\begin{equation}
\label{eq:edac}
e_i = e_{collision,i} + e_{fairness,i}.
\end{equation}

The first term of (\ref{eq:edac}) ensures that the collision probability in the network is driven to the target value:
\begin{equation}
e_{collision,i} = p_{obs,i} - p_{opt},
\end{equation}
where $p_{obs,i}$ denotes the collision probability as measured by station $i$. When the collision probability observed by station $i$ is larger than the target value, the above term yields a positive error that increases the $CW$ of station $i$, thereby reducing the collision probability.

The second term of (\ref{eq:edac}) is computed as
\begin{equation}
\label{eq:efair}
e_{fairness,i} = p_{obs,i} - p_{own,i},
\end{equation}
where $p_{own,i}$ is the collision probability experienced by station $i$. The purpose of this second component of $e_i$ is to drive the $CW$ of all stations to the same value. Indeed, the higher the $CW_{min}$, the lower the number of collisions caused, and thereby, the lower the observed collision probability $p_{obs,i}$ is. Therefore, a station will increase its $CW_{min}$ if it experiences less collisions than the others.

\revs{
With the above, the error signal at the input of the controller can be expressed as:
\begin{equation}
\label{eq:e_dac}
e_i = 2\cdot p_{obs,i} - p_{own,i} - p_{opt}.
\end{equation}
}

\begin{algorithm}[!t]
 \caption{Distributed Adaptive Control algorithm.}
 \label{alg:dac}
 \begin{algorithmic}[1]
	\While{true}
		\Repeat
			\If{new frame sniffed}
					\State retrieve retry flag and 
					\State increment $R_0$ or $R_1$ accordingly
			\EndIf
		\Until{beacon received}
		\State compute $p_{obs,i}$ using (\ref{eq:p_others}) 	
		\State fetch $T$ and $F$ from driver stats
		\State compute $p_{own,i}$ using (\ref{eq:pown})
		\State $e[t] = 2\cdot p_{obs,i}[t]-p_{own,i}[t]-p_{opt}$	
		\State $CW_{min}[t] = CW_{min}[t - 1] + K_P \cdot e[t] + $
		\State \hspace{4.2em} $ + (K_I - K_P) \cdot e[t - 1]$
		\State update the local $CW$ configuration
	\EndWhile
\end{algorithmic}
\end{algorithm}

To compute the error signal, each station needs to measure $p_{obs,i}$ and $p_{own,i}$. The former is computed as $p_{obs}$ in CAC. For the computation of $p_{own,i}$, we rely on the following statistics which are readily available from wireless cards: the number of successful transmission attempts $T$ and the number of failed attempts $F$. With these statistics, $p_{own,i}$ is computed as:
\begin{equation} 
\label{eq:pown}
p_{own,i} = \frac{F}{F+T}.
\end{equation}

Each station will estimate $p_{obs,i}$ and $p_{own,i}$ and compute the error signal $e_i$, which is provided to the PI controller for the computation of the new $CW_{min,i}$. Like in CAC, we choose to trigger an update of the $CW_{min,i}$ every beacon interval, as this is compatible with existing 802.11 hardware, which is able to update the EDCA configuration at the beacon frequency. 

Although the analysis of DAC, based on multivariable control theory, significantly differs from the analysis of CAC, based on standard control theory, the $\{K_P, K_I\}$ parameters that each station uses are the same ones of (\ref{eq:ctrl_params}), as proved in \cite{patras10tmc}. The DAC operation is summarized in Algorithm~\ref{alg:dac}.

\revs{Finally, it is important to notice that both approaches do not require to estimate the number of active stations nor additional signaling among nodes, thus scale well with the network size. However, DAC fits more naturally the ad-hoc mode of operation where there isn't a single entity responsible for maintaining node synchronization and propagating the MAC parameters through beacons. On the other hand, we will show that by having a global view of the WLAN conditions, CAC can ensure better fairness among stations. Therefore, a tradeoff between throughput fairness and algorithm operation paradigm is made when choosing one of the two approaches in a practical deployment.}

\section{Implementation Details}
\label{sec:implementation}

A major advantage of CAC and DAC is that they are based on functionalities already available in IEEE 802.11 devices, and therefore can be implemented with \revs{commercial off-the-shelf (COTS)} hardware. \revs{We address the WLAN operation with nodes employing a single access category, namely best-effort, as we observe applications request service differentiation at the IP layer only occasionally through setting the ToS field. As such, the MAC layer is unable to classify the received packets, which are instead assigned to a single queue.\footnote{In \cite{tomccap} we present a simple prototype that optimizes the performance of video in the presence of data traffic while employing different access-categories.}} In what follows we describe the hardware used in our deployment and the implementation of the functionality required by CAC and DAC. 

\subsection{Implementation Overview}
We have implemented our algorithms using Soekris net4826-48 devices.\footnote{{\ttfamily \url{http://www.soekris.com/}}} These are low-power, low-costs computers equipped with 233MHz AMD Geode SC1100 CPUs, 2 Mini-PCI sockets, 128 Mbyte SDRAM and 256 Mbyte compact flash circuits for data storage. To accommodate the installation of current Linux distributions, we have extended the storage capacity of the boards with 2-GB USB drives. As wireless interfaces, we used Atheros AR5414-based 802.11a/b/g devices. 

As software platform we installed Gentoo Linux OS (kernel 2.6.24) and the popular MadWifi open-source WLAN driver\footnote{\url{http://madwifi-project.org/}} (version v0.9.4), which we modified as follows: ($i$)~\revs{since packets delivered by the applications are placed by default in the best effort queue and we noticed the driver had disabled the configuration of the MAC parameters for this access category, thus conflicting with the standard specification,} we enabled the dynamic setting of the EDCA parameters for this AC ($ii$)~we overwrote the drivers' EDCA values for the best-effort traffic with the standard recommended ones \cite{80211revised}, and ($iii$)~for the case of DAC we modified the driver to enable the stations to employ the locally computed EDCA configuration using standardized system calls (as described in Section~\ref{sec:cw_update}). The source code of the modified drivers and our implemented prototypes is available online.\footnote{\url{http://www.hamilton.ie/ppatras/#code}}

\begin{figure}%
\includegraphics[width=\columnwidth]{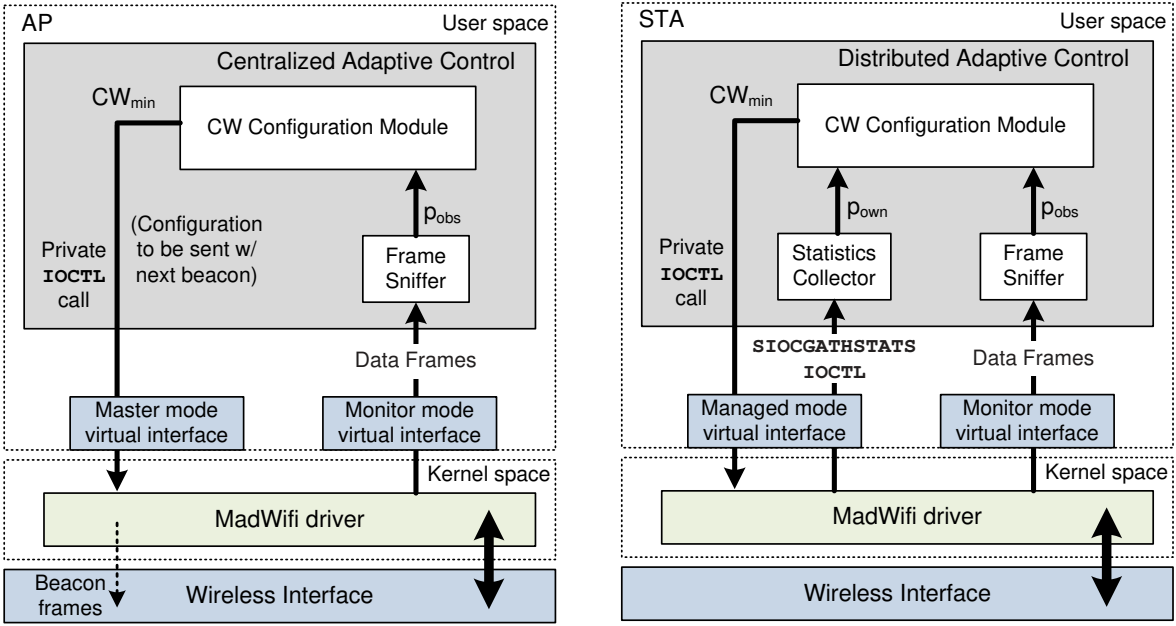}%
\caption{CAC \revs{(left)} and DAC \revs{(right)} implementations.}%
\label{fig:implementation}%
\end{figure}

Fig.~\ref{fig:implementation} illustrates the main modules of our implementation of CAC and DAC. The algorithms do not require introducing modifications to the hardware/firmware nor have tight timing constraints, and therefore they can run as user-space applications that communicate with the driver by means of {\ttfamily IOCTL} calls. We also take advantage of the ability of the MadWifi driver to support multiple virtual devices using different operation modes (master/managed/monitor) with a single physical interface. In the following we detail the implementation of the different modules. 

\subsection{Estimation of p$_{obs}$}

Both algorithms require to estimate the collision probability observed in the WLAN. For the case of CAC this is performed only at the AP and results in $p_{obs}$, while for the case of DAC this is performed independently at each station $i$ and results in $p_{obs,i}$. The estimators are computed with (\ref{eq:p_others}), which relies on observing the retry flag of the overheard frames. We next explain how these values are obtained from a practical perspective. 

To overhear frames, we utilize a virtual device operating in the so called {\ttfamily monitor} mode with promiscuous configuration. With this configuration, the device passes all traffic to user-space applications, including frames not addressed to the station. We also configure the device to pass the received frames with full IEEE 802.11 link layer headers, such that the Frame Control field of the frames (where the retry flag resides) can be examined. 

With this set-up, the algorithms open a {\ttfamily raw} socket to the driver, which enables the reception of Layer~2 frames. Through this socket, the algorithms listen for transmitted frames and process their headers in an independent thread (the ``Frame Sniffer'' module of Fig.~\ref{fig:implementation}). For every observed frame, one of the counters used in the estimation of the collision probability is incremented: $R_0$ if the retry flag was unset, $R_1$ if the retry flag was set. Every beacon interval the computation of $p_{obs}$ or $p_{obs,i}$ using~(\ref{eq:p_others}) is triggered, and then the counters are reset to zero. 

\subsection{Estimation of p$_{own}$}
\label{sec:pown}

In addition to the observed collision probability $p_{obs,i}$, the DAC algorithm requires to estimate the experienced collision probability $p_{own,i}$. We perform this computation in the ``Statistics Collector'' module of Fig.~\ref{fig:implementation} using information recorded by the wireless driver. More specifically, at the end of a beacon interval we open a communication channel with the driver instance, configured in {\ttfamily managed} mode, and perform a {\ttfamily SIOCGATHSTATS IOCTL} request. Upon this request, the driver populates an {\ttfamily ath\_stats} data structure, which contains detailed information about the transmitted and received frames since the Linux kernel has loaded the driver module. Out of the statistics retrieved, the records that are of particular interest for our implementation are: 
\begin{itemize}
  \item {\ttfamily ast\_tx\_packets}: number of unique frames sent to the transmission interface.
  \item {\ttfamily ast\_tx\_noack}: number of transmitted frames that do not require ACK.
  \item {\ttfamily ast\_tx\_longretry}: number of transmission retries of frames larger than the RTS threshold. As we do not use the RTS/CTS mechanisms, this is the total number of retransmissions.
  \item {\ttfamily ast\_tx\_xretries}: number of frames not transmitted due to exceeding the retry limit, which is set by the {\ttfamily MAX\_RETRY} parameter.
\end{itemize}

To compute $p_{own,i}$ we need to count the number of successful transmissions and the number of failed attempts. To compute the former, we subtract from the number of unique frames those that are not acknowledged (e.g., management frames) and those that were not delivered,

\begin{center}
 \emph{Successes} = {\ttfamily ast\_tx\_packets $-$ ast\_tx\_xretries $-$ ast\_tx\_noack.}
\end{center}
Similarly, to compute the number of failed attempts, out of the total number of retransmissions we do not count those retransmissions caused by frames that were eventually discarded because the {\ttfamily MAX\_RETRY} limit was reached, therefore, 
\begin{center}
 \emph{Failures} = {\ttfamily ast\_tx\_longretry $-$ ast\_tx\_xretries~$\cdot$~MAX\_RETRY.}
\end{center}
With the above, the terms $F$ and $T$ of (\ref{eq:pown}) used to estimate $p_{own,i}$ are computed as
\begin{center}
\emph{
F[t] = Failures[t] - Failures[t-1], \\
T[t] = Successes[t]- Successes[t-1],}
\end{center}
where $t$ denotes the time of the current beacon interval and $t-1$ the previous one. \revs{Notice that, in the above we do not take the differences of the observed values within two consecutive intervals, but rather the rather the difference between the total number of failures (respectively successes) counted by the driver at the end of the previous and respectively of the current interval, as these counters cannot be reset by the user and maintain absolute values accumulated since the driver module has been loaded by the kernel.}

\subsection{Contention Window Update}
\label{sec:cw_update}
With the estimated collision probabilities, CAC and DAC compute the error signal at the end of a beacon interval according to (\ref{eq:e_cac}) and (\ref{eq:edac}), respectively. Depending on this value, the PI controller triggers an update of the $CW_{min}$ to be used in the next beacon interval $t$, according to the following expression:
\[CW_{min}[t] = CW_{min}[t - 1] + K_P \cdot e[t] + (K_I - K_P) \cdot e[t - 1].\]

To ensure a safeguard against too large and too small $CW_{min}$ values we impose lower and upper bounds for the $CW_{min}$. We set these bounds to the default $CW_{min}^{DCF}$ and $CW_{max}^{DCF}$ values specified by the standard, which are 16 and 1024, respectively, for IEEE 802.11a~\cite{80211A}.

\begin{figure}[!t]%
\includegraphics[width=\columnwidth]{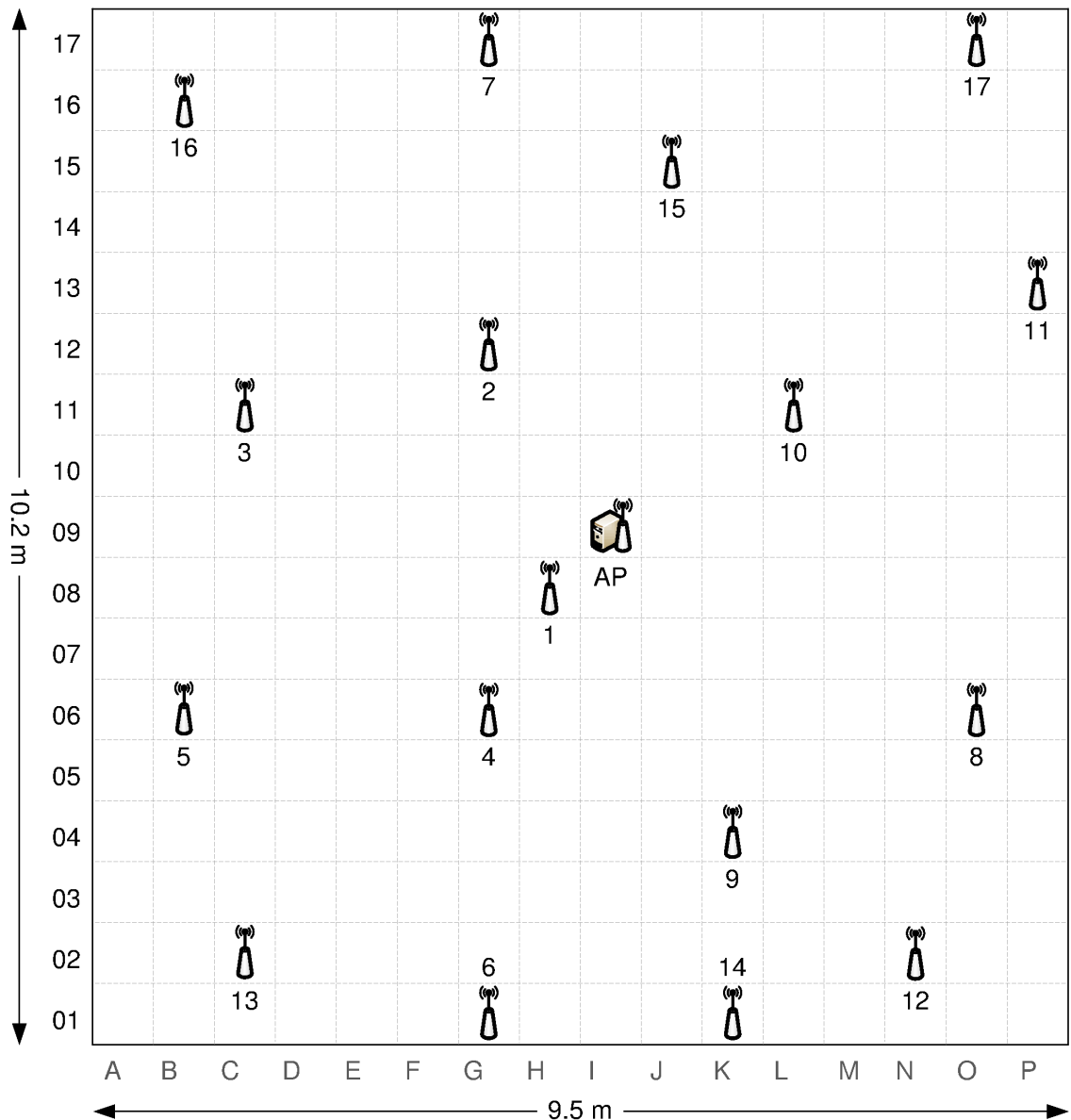}%
\caption{Deployed testbed.}%
\label{fig:testbed}%
\end{figure}

The algorithms assume that the $CW_{min}$ can take any integer value in the $[16,1024]$ range. However, with our devices only integer powers of 2 are supported (i.e., $CW_{min} \in \{16, 32, \ldots, 1024\}$). Therefore, the value actually used is obtained as:
\begin{center}
{\ttfamily CW[t] = pow(2,rint(log$_2$(CW$_{\min}$[t])))}. 	
\end{center} 
where {\ttfamily rint(x)} is a function that returns the integer value nearest to {\ttfamily x}. 

To commit the computed {\ttfamily CW} configuration, first we retrieve the list of private {\ttfamily IOCTL}s supported by the device to search for the call that sets the $CW_{min}$. Once this call has been identified, we prepare an {\ttfamily iwreq} data structure with the following information: the interface name, the base-2 exponent of the {\ttfamily CW} computed as above, the access category index as defined by the standard ({\ttfamily 0} for Best Effort) and an additional parameter that identifies if the value is intended to be used locally or propagated. For the case of DAC this value is set to 0, as the {\ttfamily CW} is only intended to the local card, while for the case of CAC is set to 1, thereby requesting the driver to broadcast the new {\ttfamily CW} within the EDCA Parameter Set element of the next scheduled beacon frame. 

\begin{figure}[t]%
\includegraphics[width=\columnwidth]{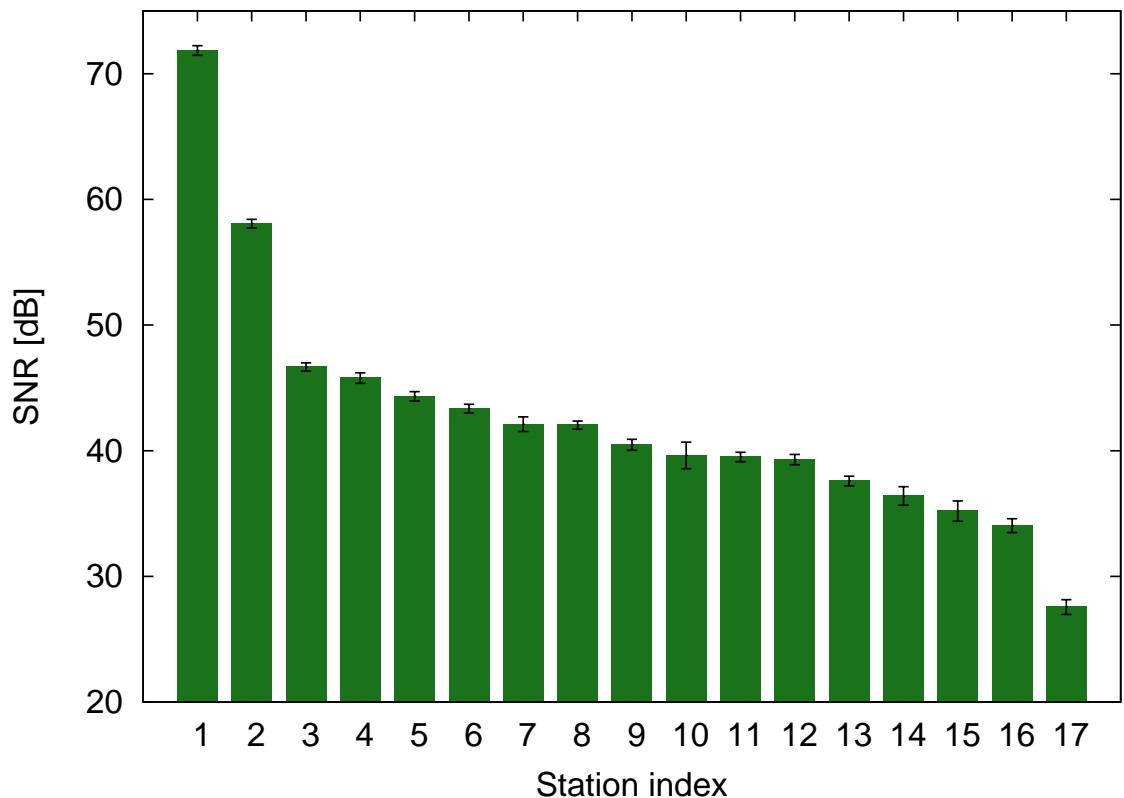}%
\caption{SNR of the links between each node and the Access Point.}%
\label{fig:snr}%
\end{figure}

\section{Testbed Description and Validation of the Implementation}
\label{sec:testbed}

In this section we first describe our testbed. Then we analyze the link qualities between each node and the AP and show that our set-up is able to mimic a realistic deployment with significant differences in terms of SNR. Finally, we confirm that, despite the constraints imposed by the devices and the realistic radio conditions, both CAC and DAC are able to drive the WLAN to a stable point of operation. 

\subsection{Testbed Description}

Our testbed is located in the Torres Quevedo building at University Carlos III de Madrid. It consists of 18 devices deployed under the raised floor, a placement that provides physical protection as well as radio shielding to some extent (see \cite{serrano10eurasip}). 

Fig.~\ref{fig:testbed} illustrates the location of the nodes. We placed one node (denoted as AP) towards the center of the testbed, thus following the placement of an Access Point in a realistic deployment, while the other stations (numbered from 1 to 17 in \revs{decreasing order of their links' SNR values, which we measure as explained in what follows}) are distributed at different distances from \revs{the AP}. \revs{We argue that the size of our deployment is representative for several popular scenarios, e.g. caf\'es, offices, conference rooms, etc., whereby nodes are mostly static, while having reasonable link qualities to the AP but experiencing different SNR values.\footnote{In static deployments SNR values are expected to have only small variations, as our measurements confirm; hence, rate control algorithms are seldom necessary.} Although we consider a single AP topology, we note that under the considered scenarios careful channel allocation is performed to avoid inter-AP interference when more than one AP is deployed for accommodating more clients. As such, our algorithms can be independently executed within each deployed cell.}

All nodes are equipped with 5~dBi omnidirectional antennas and are configured to operate on channel 64 (5.32 GHz) of IEEE 802.11a standard \cite{80211A}, where no other WLANs were detected. All nodes use the 16-QAM modulation and coding scheme, which provides 24~Mbps channel bit rate, as calibration measurements showed that this was the highest rate achievable by the node with the worst link to the AP (node 17). Additionally, we disabled the RTS/CTS, rate adaptation, turbo, fast frame, bursting and unscheduled automatic power save delivery functionality, as well as the antenna diversity scheme for transmission/reception. 

Unless otherwise specified, all nodes use the same transmission power level of 17~dBm. Given the node placement of Fig.~\ref{fig:testbed}, this setting results in dissimilar link qualities between each station and the AP (e.g., node 1 is \revs{very close to the AP, while e.g. node 10 is further away, which will affect the perceived SNR}). To confirm this link heterogeneity, we designed the following experiment. For a given node, we ran a 10-second {\ttfamily ping} test between the station and the AP, recording the SNR values of the received frames as obtained by the {\ttfamily wireshark} packet analyzer\footnote{\url{http://wireshark.org}} from the {\ttfamily radiotap} header.\footnote{With the {\ttfamily radiotap} option, the driver provides additional information about received frames to user-space applications, including the signal-to-noise ratio.} This test was performed on a node-by-node basis, and repeated for 18 hours. The average and standard deviation of the SNR for each link are shown in Fig.~\ref{fig:snr}. 


\subsection{Validation of the Algorithms}

Our first set of experiments aims at confirming that the good operation properties of CAC and DAC, obtained analytically and via simulations in \cite{patras09monet,patras10tmc}, are also achieved in a real testbed. Specifically, we want to confirm that the use of the algorithms results in stable behavior despite the described hardware/software limitations and the impairments introduced by the channel conditions, and also assess their resource consumption in terms of CPU and memory usage.

\begin{figure}[t!]
\hspace{-0.3cm}
\begin{minipage}[b]{0.5\linewidth}
\centering
\includegraphics[height=\columnwidth,angle=270]{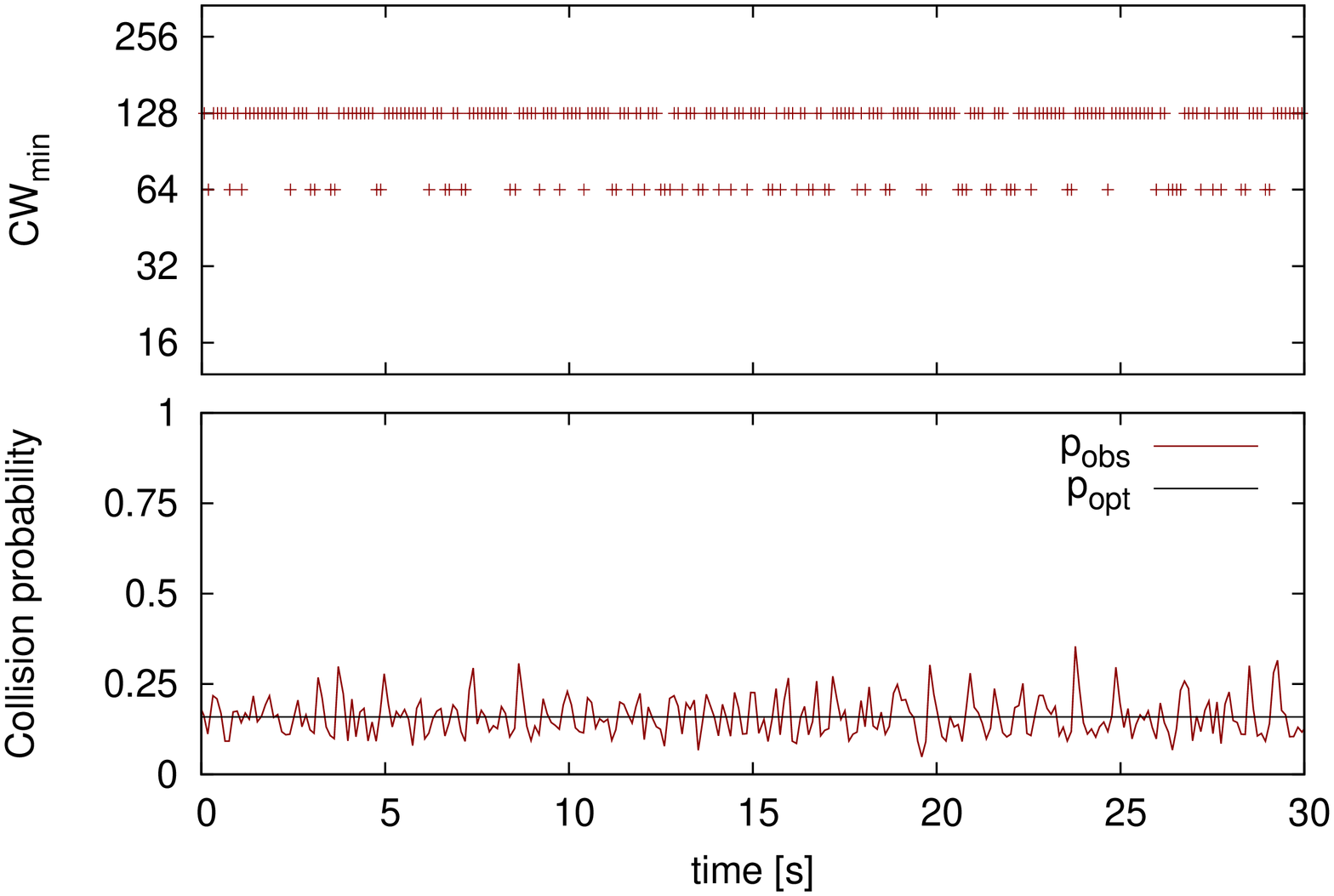}%
\caption{Announced $CW_{min}$ and observed collision probability with CAC.}%
\vspace{1.5em}
\label{fig:cac_operation}%
\end{minipage}
\hspace{0.5cm}
\begin{minipage}[b]{0.5\linewidth}
\centering
\includegraphics[height=\columnwidth,angle=270]{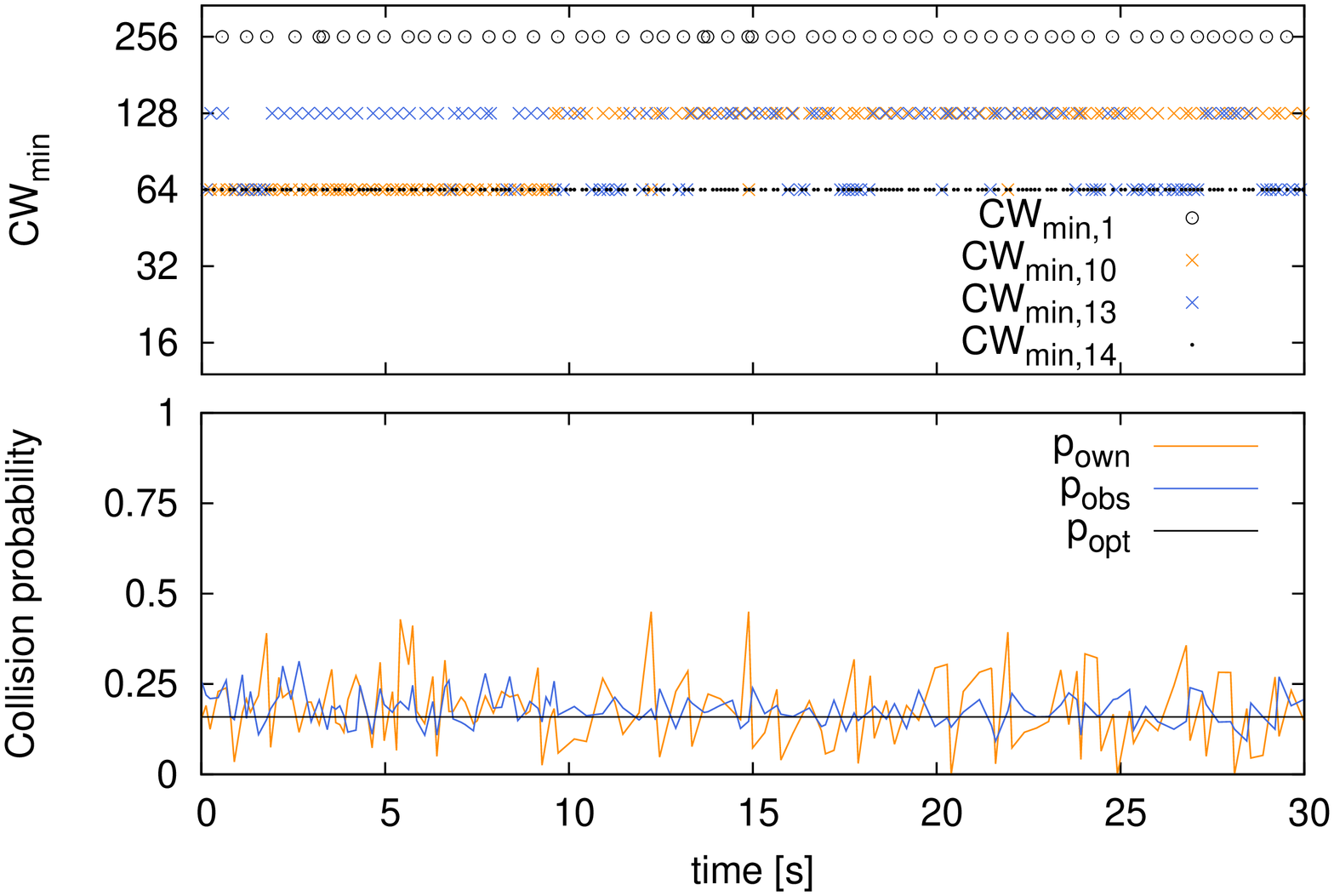}%
\caption{$CW_{min}$ used by four selected nodes \revs{(1, 10, 13 and 14)} and the estimated $p_{obs}$ and $p_{own}$ \revs{for node 10} with DAC.}%
\label{fig:dac_operation}%
\end{minipage}
\end{figure}

\vspace{0.25em}
{\bf Point of operation.} We consider a scenario with $N=10$ stations, \revs{randomly selected out of the total available in our testbed, which corresponds to a network with heterogeneous link qualities. Stations are} constantly backlogged with 1500-Byte UDP frames, which they send to the AP utilizing {\ttfamily iperf}.\footnote{{\ttfamily \url{http://sourceforge.net/projects/iperf/}}} For the case of the centralized algorithm (CAC) we log its key variables, namely, the {\ttfamily CW} announced in beacon frames and the observed collision probability $p_{obs}$. Both are obtained every 100~ms and depicted in Fig.~\ref{fig:cac_operation}.

As the figure shows, CAC drives the WLAN to the desired point of operation. Indeed, the announced {\ttfamily CW} oscillates between the two \revs{power of 2} values closest to the optimal $CW_{min}$ \revs{(i.e., 64 and 128)}, while $p_{obs}$ fluctuates stably around the desired $p_{opt}$ given by (\ref{eq:p_opt}). We conclude that, despite the hardware limitations imposed on the values of {\ttfamily CW} and the channel impairments, CAC is able to drive the WLAN to the desired point of operation. 

Next we validate the operation of the distributed algorithm (DAC). We consider the same scenario as before, logging the key parameters of the algorithm at each station, namely $CW_{min,i}$, $p_{own,i}$ and $p_{obs,i}$. In Fig.~\ref{fig:dac_operation} we depict in the upper subplot the evolution of the $CW_{min}$ used by four representative nodes. \revs{Namely, we select out of the 10 nodes used in this experiment the ones with the highest (node 1) and respectively the poorest link quality (node 14), as well as two stations with similar link qualities, (nodes 10 and 13).} Additionally, in the lower subplot we show the collision probabilities estimated by node~10 ($p_{obs,10}$ and $p_{own,10}$). 

From the two subplots we see that DAC also drives the average collision probability in the WLAN to the desired value. However, there is a key difference as compared to the previous case: while with CAC all stations use the same $CW_{min}$ value, with DAC they operate at different average $CW_{min}$. Indeed, the four stations considered in the experiment use average $CW_{min}$ values of 300, 92, 92 and 64, respectively. As we will explain in Section~\ref{sec:snr}, this behavior is caused by the relative differences in link qualities, combined with the inability of the wireless interface to identify the reasons for a packet loss.

\vspace{0.25em}
{\bf Resource consumption.} In addition to analyzing the performance of CAC and DAC, it is also important to assess their resource consumption. For this purpose, we analyzed the CPU and memory usage of the algorithms utilizing the {\ttfamily top} Linux application, which provides a dynamic real-time view of a running system. With this tool, we recorded the used shares of the CPU time and available physical memory with a frequency of 1 sample per second and computed the average usage. CAC, which runs exclusively at the AP, demands on average 14\% of the CPU time and only 0.8\% of the physical memory. For the case of DAC, which runs at every station, the average CPU time consumption is 17\%, while the physical memory consumption is 0.9\%. Given the low speed of the nodes' CPU (233 MHz) and their reduced physical memory (128 MB), these results show that both CAC and DAC are suitable for commercial deployments. \revs{Further, we argue that the main source for resource consumption is the packet capture routine. To verify this hypothesis, we compared the requirements of the two prototypes with the CPU and memory usage of the popular packet sniffer {\ttfamily tcpdump}, running on the same device type. Measurements revealed that {\ttfamily tcpdump} consumes 28\% of the CPU time, while its memory usage reaches 4.3\%. We conclude that our implementations do not pose stringent requirements to commodity hardware.}

\begin{figure}[!t]%
\includegraphics[height=\columnwidth, angle=270]{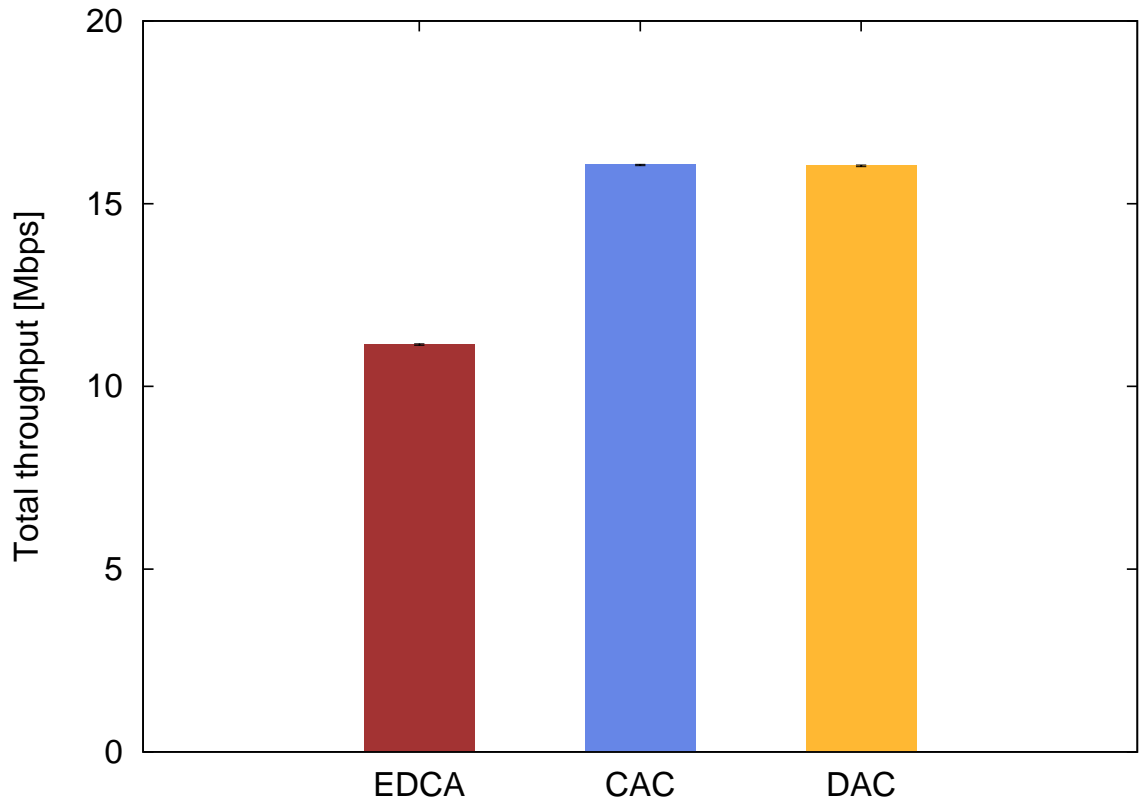}%
\caption{Total throughput with UDP traffic. }%
\label{fig:udp_total}%
\end{figure}

\section{Performance Evaluation}
\label{sec:performance}

We next assess the performance of the algorithms under a large number of different scenarios and evaluate their improvements over the default EDCA configuration, which we use as a benchmark. Each considered experiment runs for 2~minutes and is repeated 10 times to obtain average values of the measured metrics with good statistical significance. 

\revs{We evaluated the behavior of the proposed algorithms with both unidirectional (UDP) and bidirectional (TCP) traffic, under diverse conditions in terms of link qualities and number of active nodes. Unless otherwise stated, we assume nodes have similar traffic demands and are constantly backlogged, thus the network is mostly saturated.}


\subsection{UDP Throughput}

We first measure the achievable throughput between the nodes and the AP when all the stations are transmitting UDP traffic at the same time. Fig.~\ref{fig:udp_total} plots the average and standard deviation of the total throughput obtained with each mechanism. We observe that the EDCA default configuration achieves around 11~Mbps, while the use of DAC and CAC results in a performance gain of approximately 45\%. Therefore, we confirm that both approaches, by properly adapting the $CW$ configuration to the number of contending stations, achieve a much higher efficiency. 

\begin{figure}[t]%
\includegraphics[height=\columnwidth,angle=270]{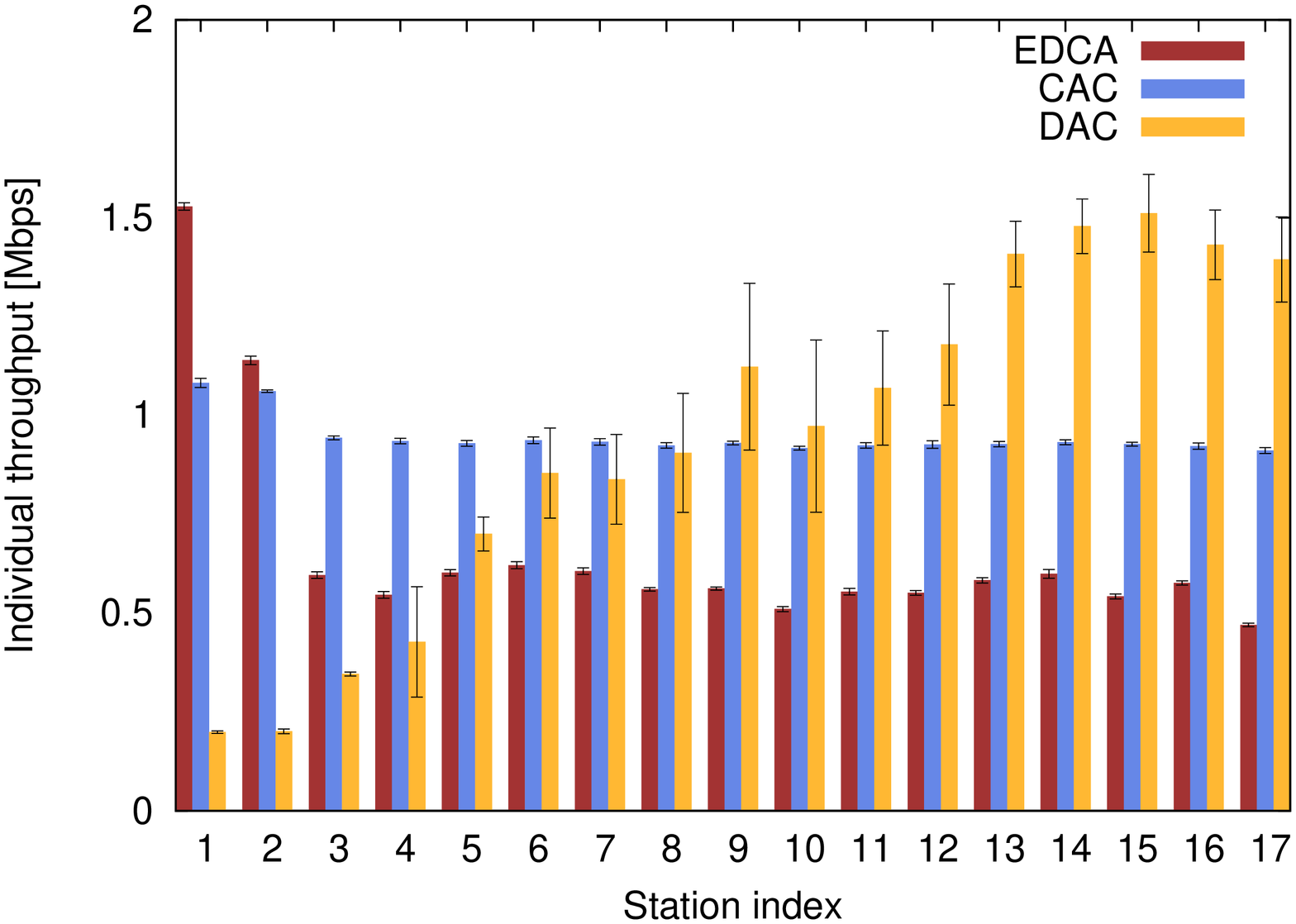}%
\caption{Throughput per station with UDP traffic.\vspace{-0.5em}}%
\label{fig:udp_throughput_stations}%
\end{figure}

To further examine the performance of the algorithms we plot the per-station throughput in Fig.~\ref{fig:udp_throughput_stations}. According to the figure, the use of the EDCA recommended values not only provides the lowest overall throughput figures, but also fails to provide a fair sharing of the available bandwidth. Indeed, it can be seen that, e.g., the node with the best link quality to the AP (node 3) achieves more than three times the throughput obtained by the station with the poorest link (node 15).

While DAC provides a larger total throughput than EDCA, it does not improve the level of fairness. Actually, it results in a somehow \emph{opposite} performance as the one obtained with EDCA: stations that obtained a relatively large bandwidth with EDCA (e.g., nodes 3, 6) now obtain a relatively small bandwidth with DAC. The use of CAC, on the other hand, provides the best performance both in terms of total throughput and fairness, as it provides all stations with very similar throughput values. 

To quantify the throughput fairness achieved by the considered mechanisms we compute the Jain's fairness index (JFI)~\cite{jain84}. The resulting JFI values are 0.865, 0.997 and 0.817 for the case of EDCA, CAC and DAC, respectively. These figures confirm the good fairness properties of CAC, and shows that DAC and EDCA suffer from a higher level of unfairness, a result that we analyze next. 

\subsection{Impact of SNR on Throughput}
\label{sec:snr}

\begin{figure}%
\includegraphics[height=0.9\columnwidth, angle=270]{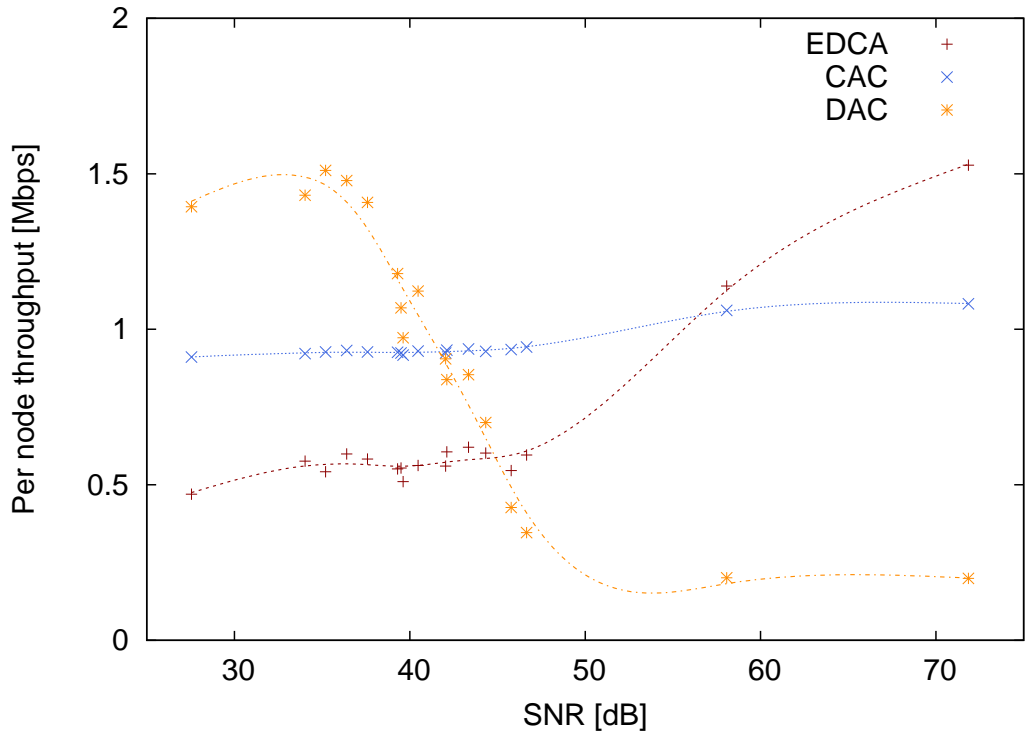}%
\caption{\revs{Per node throughput obtained vs. SNR.}}%
\label{fig:throughput_snr}%
\end{figure}

We have seen that link quality affects throughput distribution, in particular for EDCA and DAC. To analyze this impact, we plot in Fig.~\ref{fig:throughput_snr} the average UDP throughput per station vs. the SNR of the link between the station and the AP. \revs{Note that, in this analysis we do not modify stations' SNR by tuning their transmission power, but instead we investigate how the SNR dissimilarity that arises due to their relative position to the AP affects the throughput performance of CAC and DAC as compared to EDCA.} For ease of visualization we also plot \emph{natural smoothing splines} over the data points. 

From the figure we observe that: ($i$)~for EDCA there is a noticeable and positive correlation between SNR and throughput; ($ii$)~for CAC, performance is not much affected by SNR dissimilarities, as significantly better link qualities result in very small throughput improvements; ($iii$)~for DAC there is a large and negative correlation between SNR and throughput, with small differences in terms of SNR causing large differences in terms of throughput.  

For the case of EDCA, the positive correlation is caused by the \emph{capture effect} \revs{\cite{hadzi02}}. With this effect, in case of a collision the receiver can decode the packet with the higher SNR. As a result, stations with better link quality obtain higher throughput. In contrast, the use of CAC reduces the number of collisions in the WLAN, and therefore the impact of the capture effect is significantly reduced. 


For the case of DAC, the negative correlation is also driven by the capture effect as follows. Nodes with high capture probability will experience smaller collision rates than the others, and therefore will have $p_{own,i}$ smaller than $p_{obs,i}$. This will cause a positive error signal according to the $e_{fairness,i}$ term in (\ref{eq:efair}), which will result in large $CW_{min}$ values. Conversely, nodes with low capture probability will experience larger $p_{own,i}$ values and smaller $p_{obs,i}$ ones, and therefore will have smaller $CW_{min}$ configurations. In this way, capturing nodes will transmit less often and therefore will obtain low throughput figures, while the other nodes will transmit more often and experience a higher throughput. Additional experiments with different transmission power settings, not reported due to space constraints, confirmed that a careful \emph{equalization} of the link qualities is able to restore fairness to some extent. 



\begin{figure}%
\includegraphics[height=0.9\columnwidth,angle=270]{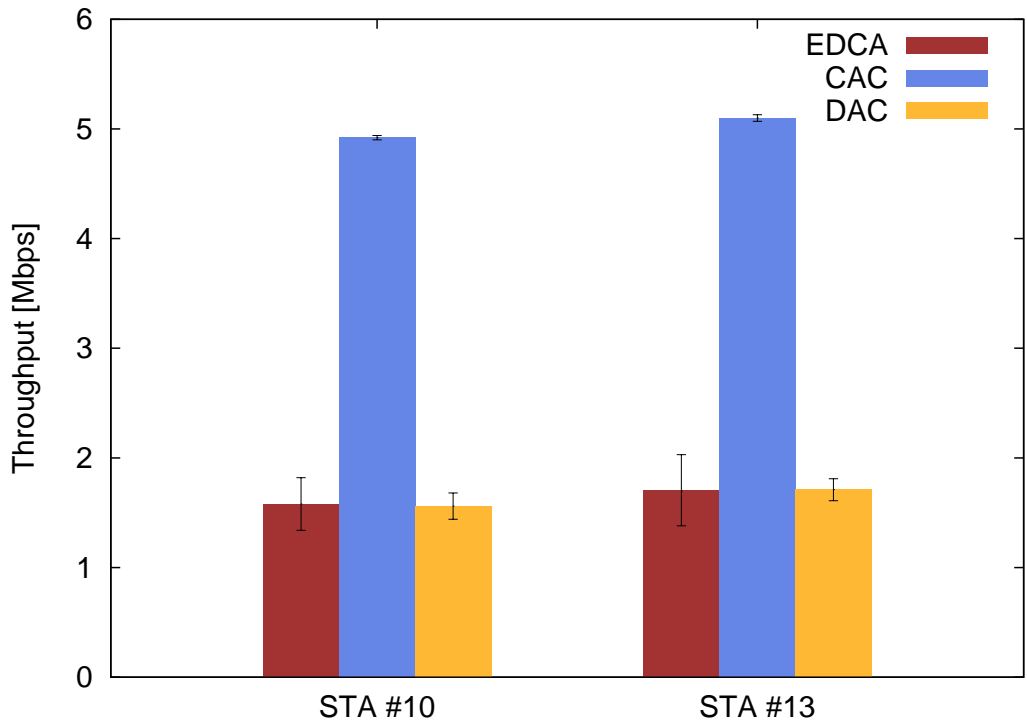}%
\caption{Performance with hidden nodes.}%
\label{fig:hidden}%
\end{figure}

\subsection{Hidden Nodes Scenario}
\label{sec:hidden}

Our adaptive algorithms have been designed for scenarios where all stations are in radio range of each other and coordinate their transmissions by means of carrier sensing. However, in real deployments hidden nodes may be present, and therefore we want to investigate their behavior under such circumstances. 

To this aim, we ran extensive measurements, selecting different topologies and different transmission power settings, to determine the most pathological scenario. This is obtained when node 1 acts as AP, and nodes 10 and 13 act as stations, using a transmission power level of 5~dBm. With this setting, each EDCA station transmitting in isolation (i.e., with the other station silent) obtains about 16.3~Mbps of UDP throughput, while if both stations transmit simultaneously the throughput of each one drops to 1.6~Mbps. Thereby we managed to reproduce a hidden node scenario.

We then repeated the experiment with CAC and DAC, and obtained the results depicted in Fig.~\ref{fig:hidden}. We observe that the use of DAC does not improve performance over EDCA. In contrast, CAC provides a dramatic throughput increase, i.e., more than three times the throughput attained with the other mechanisms. We conclude that CAC detects the large collision rate and commands hidden nodes to be less aggressive by announcing a higher $CW_{min}$, which lessens (but does not eliminate) the hidden node problem. On the other hand, a station running DAC is not able to overhear MAC (re-)transmissions from hidden nodes, and hence cannot correctly estimate the collision probability in the network.

\subsection{Impact Network Size}

\begin{figure}%
\includegraphics[height=0.9\columnwidth, angle=270]{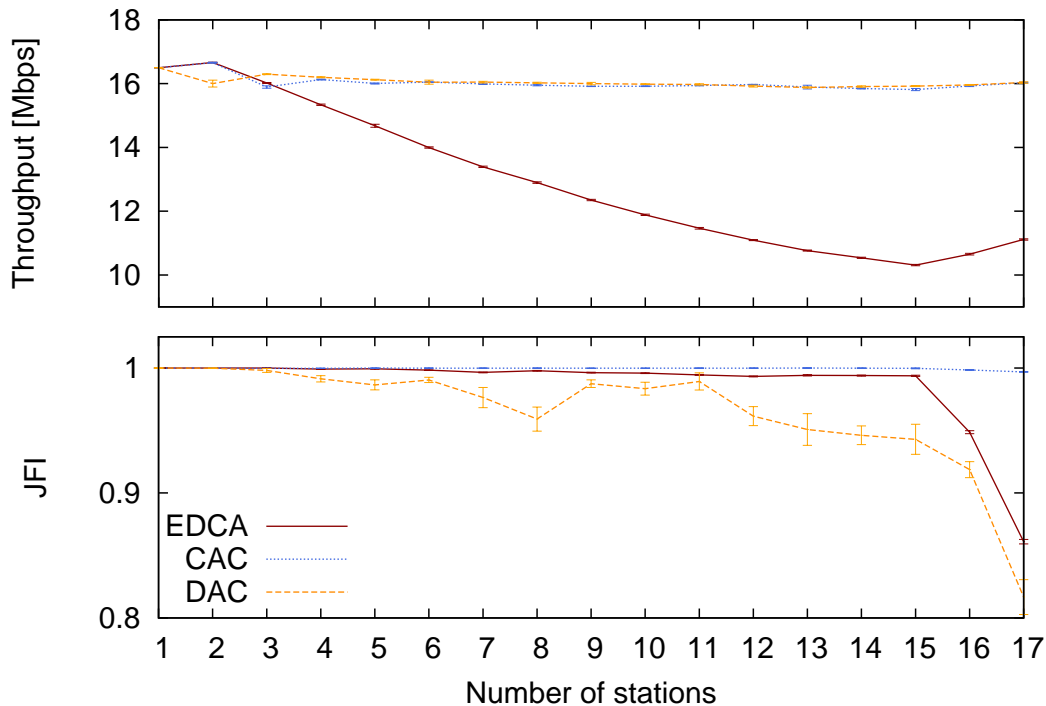}%
\caption{Total throughput and fairness for different number of stations.}%
\label{fig:throughput_n}%
\end{figure}

We next evaluate the performance of the algorithms as a function of the number of stations. To this aim, we measure the total throughput and JFI for an increasing number of contending nodes, adding new stations in ascending order of their link qualitiy. We plot the obtained results in Fig.~\ref{fig:throughput_n}.

We observe that for both DAC and CAC the total throughput performance is practically flat, regardless of the number of stations. This result confirms that both approaches are able to adapt the $CW$ to the number of stations present in the WLAN.

For the case of EDCA, performance degrades with the number of stations, which is the expected result from the use of a fixed set of (relatively small) contention parameters. However, for $N > 15$ the total throughput performance slightly grows again, a behavior caused by the capture effect as the last nodes to be added in our experiments are the ones experiencing better link qualities (nodes 2 and 1). This is confirmed by the fairness values, as for $N > 15$ there is a drop in the JFI for the case of EDCA. JFI values also confirm that DAC is more sensitive to heterogeneous link conditions, as its performance noticeably degrades with $N$. In contrast, with CAC the fairness index is practically constant for all $N$ values. 

\vspace{0.5em}
\subsection{TCP Throughput}

We next evaluate performance in scenarios in which stations use TCP. We start by evaluating the throughput and fairness performance when all stations are constantly backlogged sending TCP traffic to the AP, replicating \emph{bulky} FTP transfers. Note that this scenario is substantially different from the ones considered in the previous subsections, as TCP congestion control\footnote{The Linux distribution used in our deployment executes the TCP CUBIC variant \cite{ha08}.} introduces a ``closed loop'' that can lead to extreme unfairness conditions and even starvation \cite{gurewitz}. \revs{Precisely, we expect nodes that use a more aggressive CW configuration to leave less channel access opportunities for the more conservative ones, thus causing those to reach congestion much faster and reset their TCP windows more frequently. This behavior will exacerbate the difference in the throughput values they obtain and impact the fairness performance more severely.}

\begin{figure}[!t]%
\includegraphics[height=\columnwidth, angle=270]{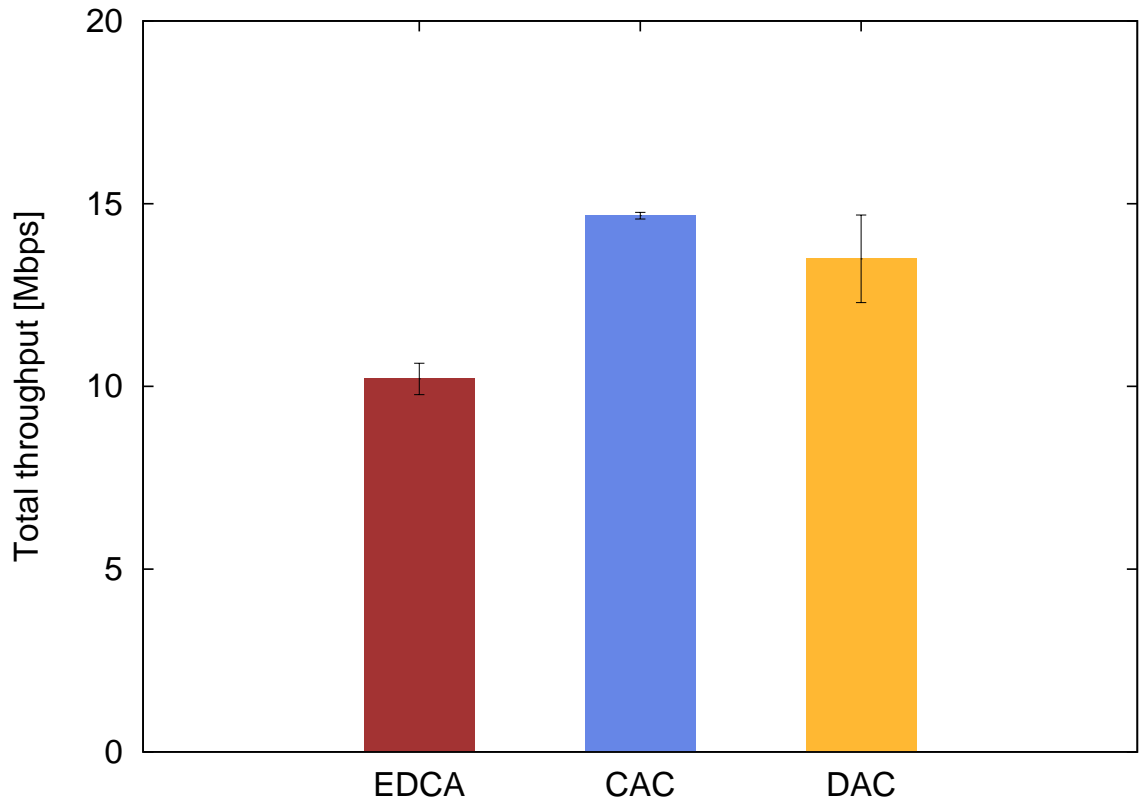}%
\caption{Total throughput of FTP-like traffic. }%
\label{fig:tcp_total}%
\end{figure}

We plot in Fig.~\ref{fig:tcp_total} the total throughput values for the three mechanisms. According to the results, both CAC and DAC significantly outperform EDCA, improving throughput by 50\% and 40\%, respectively.

\begin{figure}[t]%
\includegraphics[height=\columnwidth, angle=270]{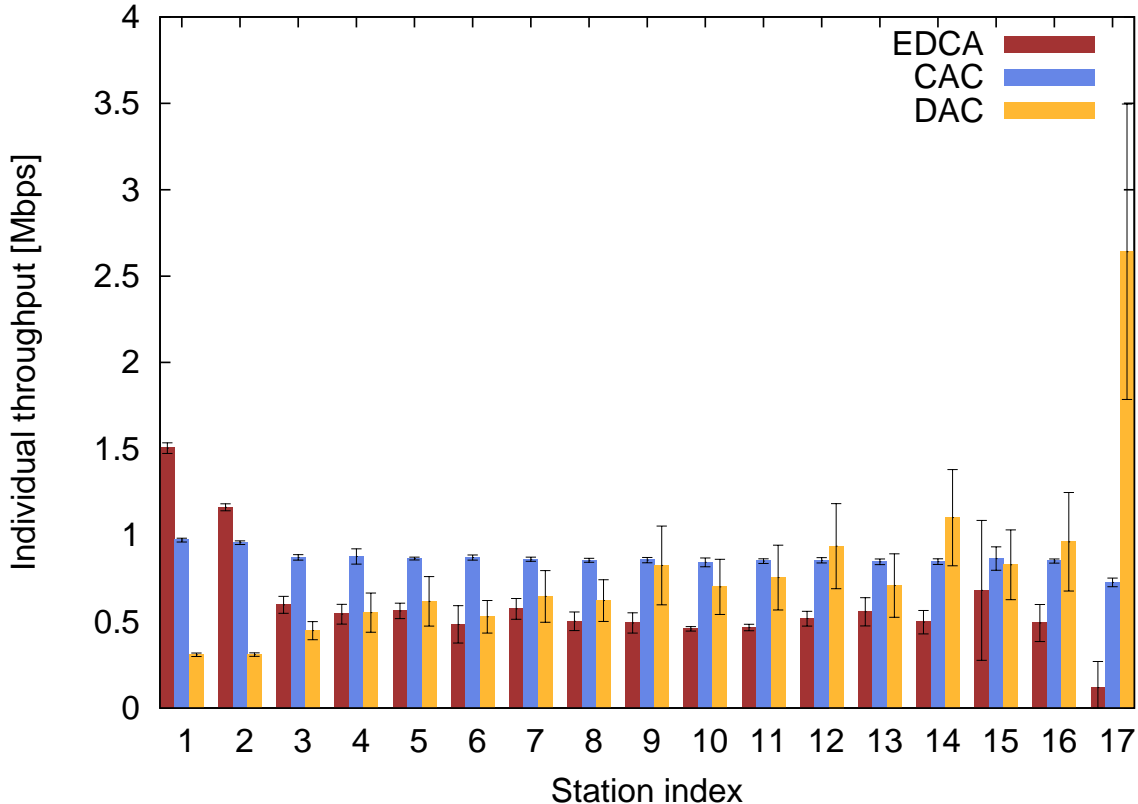}%
\caption{Throughput per station with FTP-like traffic.}%
\label{fig:tcp_individual}%
\end{figure}

\begin{figure}[p]
\centering
\includegraphics[width=0.45\linewidth, angle=270]{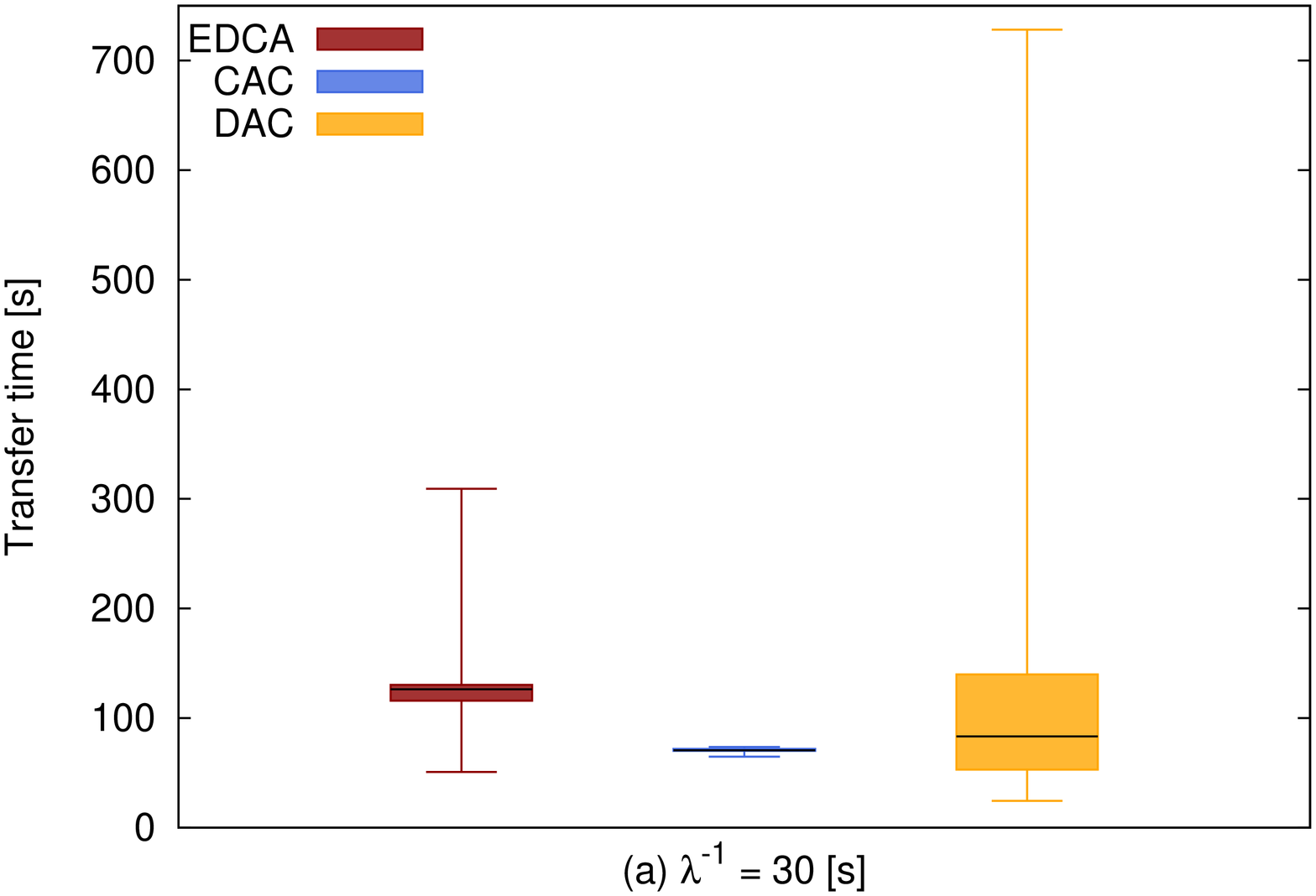}
\includegraphics[width=0.45\linewidth, angle=270]{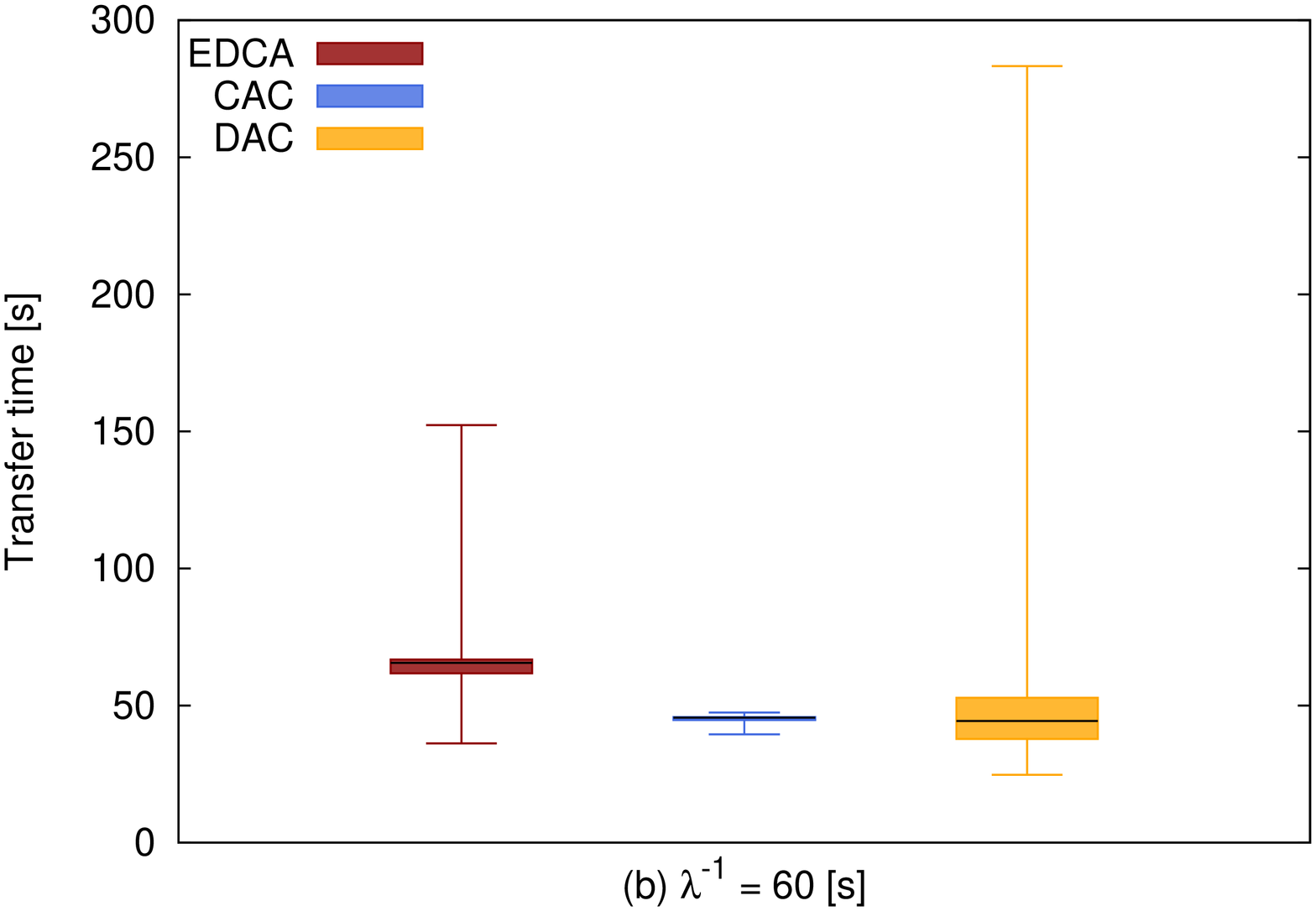}
\includegraphics[width=0.45\linewidth, angle=270]{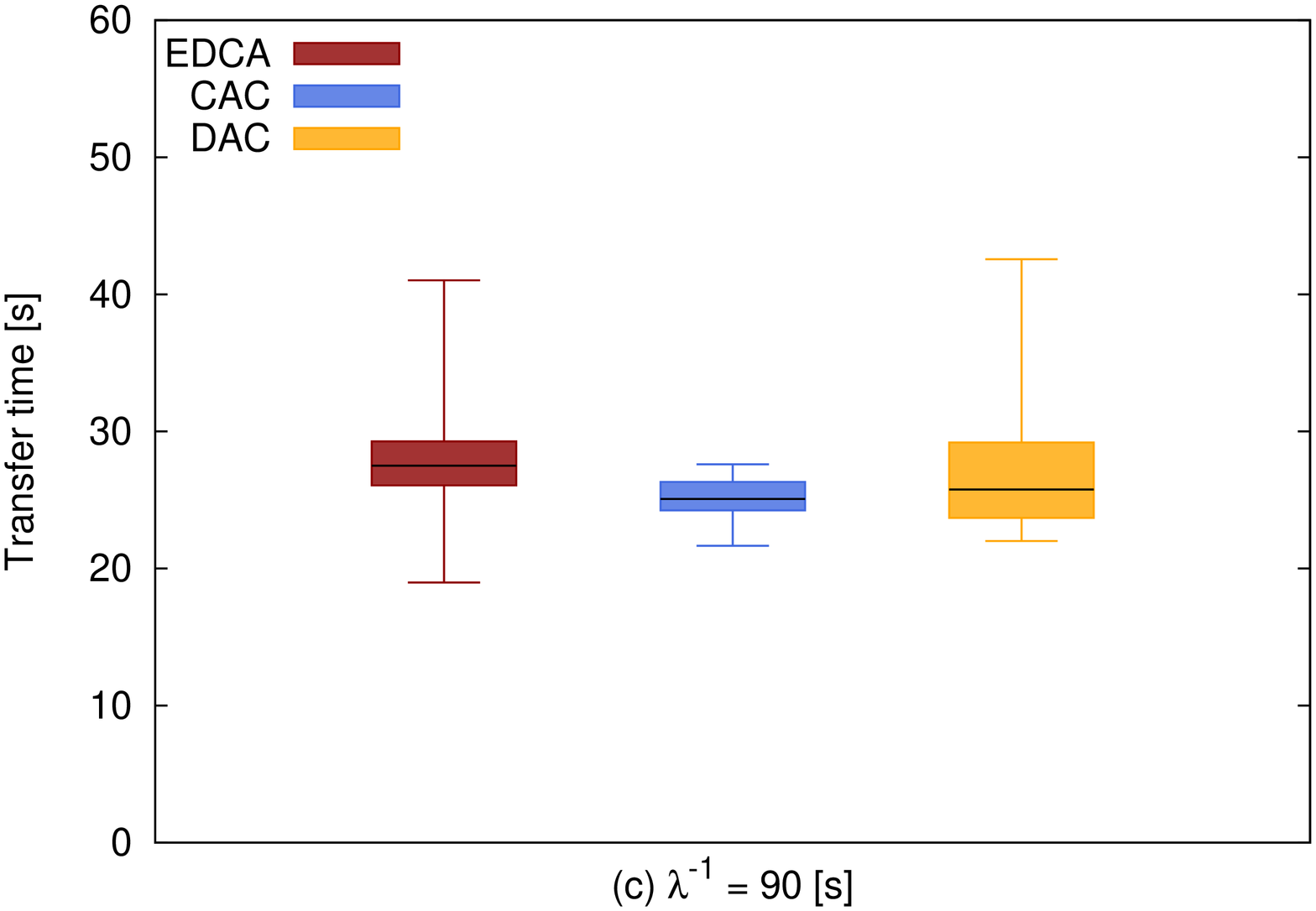}
\caption{TCP delay performance.}%
\label{fig:delay_individual}%
\end{figure}

The per-station throughput distribution is depicted in Fig.~\ref{fig:tcp_individual}. With EDCA, the node with the poorest link quality (node 17) suffers from a large performance degradation, this being worse than in the UDP case (see Fig.~\ref{fig:udp_total}). The use of DAC with TCP traffic also exacerbates the unevenness in the traffic distribution, with node 15 clearly outstanding among the other nodes. DAC results also present a large deviation, caused by relatively frequent TCP timeouts from nodes with weak radio link (e.g., node 17). Conversely, CAC yields a remarkably fair and stable throughput distribution. 


Like in the UDP case, we compute the JFI values for the resulting throughput distributions. In this case, the values for EDCA, CAC and DAC are 0.787, 0.996 and 0.692, respectively. We conclude that, as expected, the performance of EDCA and DAC worsens with TCP, while CAC preserves its good properties in this scenario.

\vspace{0.5em}
\subsection{TCP Transfer Delay}

We finally consider a scenario involving finite-size TCP connections. More specifically, all stations alternate periods of activity---during which a transmission of 10~MB occurs---with silent periods exponentially distributed with mean $\lambda^{-1}$ \cite{Crovella98}. We consider three different values for $\lambda$, corresponding to three different levels of activity, namely high, moderate and low. For each case we ran 1-hour experiments, logging all transfer durations and computing the per-station average delay. We use a \emph{box-and-whisker} diagram to illustrate the distribution of the average delay among nodes: we provide the median, first and third quartiles of the average delay, as well as its maximum and minimum values.

Results are depicted in Fig.~\ref{fig:delay_individual}. With $\lambda^{-1}=30~s$, which corresponds to high activity, we see that CAC provides the smallest and most uniform distribution of transfer delay among nodes, with practically no difference between the best and worst performing node. In case of EDCA, the delay shows a larger median and higher variability. However, the small distance between the first and third quartiles shows that most of the stations experience similar performance. Finally, for the case of DAC, despite the median is similar to the one of CAC, results show a much larger dispersion. \revs{This further proves that, although DAC is able to achieve its main objective of optimizing the total throughput, the fact that nodes adjust their CW based on locally observed conditions and converge to different values under dissimilar link qualities leads to some stations being more aggressive than others. Thus we can explain the large variations in the observed delay figures.}

When the traffic activity is moderate ($\lambda^{-1}=60~s$), the absolute values decrease, but the relative results are similar, i.e., CAC provides again the smallest and most uniform delays among nodes. Finally, when the activity of the nodes is low ($\lambda^{-1}=90~s$), medians are very similar but still CAC provides the most fair distribution of the transfer delays. From these experiments, we conclude that CAC also provides the best performance under dynamic traffic scenarios. 

\vspace{0.5em}
\section{Related Work}
\label{sec:related}

The scientific literature offers many examples of MAC optimization approaches. Many of them are based on a centralized entity, responsible for monitoring system performance and adapting the system parameters to current conditions. Other works focus on distributed approaches to adapt MAC parameters. Very little experimental work is available, and it is based on complex algorithms, non-standard functionality and small-sized networks. In the following we review the most significant contributions in each of these areas and describe the novelty of our work. 

\vspace{0.5em}
{\bf Centralized approaches.} A significant number of approaches exists in the literature \cite{freitag06,nafaa05,siris06,patras09monet} that use a single node to compute the set of MAC parameters to be used in the WLAN. With the exception of our CAC algorithm \cite{patras09monet}, the main drawbacks of these approaches are that they are either based on heuristics, thereby lacking analytical support for providing performance guarantees \cite{freitag06,nafaa05}, or they do not consider the dynamics of the WLAN under realistic scenarios \cite{siris06}.

\vspace{0.5em}
{\bf Distributed approaches.} Several works \cite{yang07,ni03,heusse05,AOBorig,cali00} have proposed mechanisms that independently adjust the backoff operation of each stations in the WLAN. The main disadvantages of these approaches are that they change the rules of the IEEE 802.11 standard and therefore require introducing significant hardware or firm-ware modifications.

\vspace{0.5em}
{\bf Implementation experiences.} Very few schemes to optimize WLAN performance have been developed in practice \cite{siris06,AOBimpl,idle07}. While the idea behind Idle Sense \cite{heusse05} is fairly simple, its implementation \cite{idle07} entails a significant level of complexity, introducing tight timing constraints that require programming at the firmware level. \revs{Nonetheless, the microcode that achieves the desired functionality is proprietary and thus subject to portability constraints.} Similar limitations hold for the approach of \cite{AOBimpl}, which introduces changes to the MAC protocol that require redesigning the whole NIC implementation. \revs{Further, this involves complex DSP and FPGA programming and demands non-negligible computational resources.} Finally, the work of \cite{siris06} does not propose or evaluate any adaptive algorithm to adapt the $CW$ but just evaluates the performance of static configurations. Additionally, all of these works rely on testbeds substantially smaller than ours.







\section{Conclusions}
\label{sec:conclusions}

We have prototyped with standard 802.11 devices two adaptive mechanisms that tune the contention window based on the observed network conditions. In contrast to other proposals that require complex modifications, these mechanisms rely on functionalities already supported by COTS hardware/firmware, and do not introduce any extensions to the standard 802.11 MAC. We have extensively evaluated the performance of the mechanisms in an 18-nodes testbed, considering a large variety of network conditions. With our experimental study we have identified the key limitations of the distributed scheme, inherent in realistic scenarios, and we have confirmed that the centralized mechanism significantly improves network throughput, transfer delay and fairness among stations in a broad range of circumstances, including the pathological case of hidden nodes. A major conclusion from our work is that, by simply adding a few lines of code at the AP to exploit the functionality readily available, we can achieve performance improvements of up to 50\%. We believe that the results presented herein pave the way for a widespread deployment of the centralized mechanism.

\section*{Acknowledgements}

This work has been supported by the European Community's Seventh Framework Programme (FP7-ICT-2009-5) under grant agreement n. 257263 (FLAVIA project). 

\bibliographystyle{IEEE}
\bibliography{references}

\end{document}